\documentclass{sig-alternate-05-2015}
\usepackage[utf8]{inputenc}
\usepackage{graphicx}
\usepackage[noadjust]{cite}
\usepackage{multirow}
\usepackage{amssymb}
\usepackage{mathtools}
\usepackage{listings}
\usepackage[usenames,dvipsnames,svgnames,table]{xcolor}
\usepackage[caption=false]{subfig}
\usepackage{balance}
\usepackage{epstopdf}
\usepackage[ 
  bookmarks = true,
  bookmarksopen = true,
  bookmarksnumbered = true,
  breaklinks = true,
  linktocpage,
  pagebackref = false,
  colorlinks = true,
  linkcolor = blue,
  urlcolor  = blue,
  citecolor = red,
  anchorcolor = green,
  hyperindex = true,
  hyperfigures
  ]{hyperref}
\usepackage[hyphenbreaks]{breakurl}
\usepackage{url}

\newcommand{\csubfloat}[2][]{%
  \makebox[0pt]{\subfloat[#1]{#2}}}
\newcommand{\e}[1]{{\mathbb E}\left[ #1 \right]}

\def\sharedaffiliation{\end{tabular}\newline\begin{tabular}{c}}
\newcommand{\superscript}[1]{\ensuremath{^{\textrm{#1}}}}
\def\uc3m{\superscript{*}}
\def\imdea{\superscript{\dag}}
\def\nec{\superscript{\ddag}}

\makeatletter
\def\@copyrightspace{
\@float{copyrightbox}[b]
\begin{center}
\setlength{\unitlength}{1pc}
\begin{picture}(20,3.5) 
\put(0,-0.95){\crnotice{\@toappear}}
\end{picture}
\end{center}
\end@float}
\makeatother

\begin{document}

\permission{\copyright 2016 Copyright held by the authors. This is the authors' version of the work. It is posted here for your personal use. Not for redistribution. The definitive version was published in}
\conferenceinfo{ACM MSWiM '16,}{November 13 - 17, 2016, Malta, Malta}
\isbn{978-1-4503-4502-6/16/11}
\doi{http://dx.doi.org/10.1145/2988287.2989149}

\title{Revisiting 802.11 Rate Adaptation \\from Energy Consumption's Perspective}

\numberofauthors{1}
\author{
  \alignauthor \ Iñaki Ucar\uc3m, Carlos Donato\imdea, Pablo Serrano\uc3m, Andres Garcia-Saavedra\nec, \\Arturo Azcorra\uc3m\imdea, Albert Banchs\uc3m\imdea\\
   \email{inaki.ucar@uc3m.es, \{pablo,azcorra,banchs\}@it.uc3m.es, \\carlos.donato@imdea.org, andres.garcia.saavedra@neclab.eu}
\sharedaffiliation
\begin{tabular}{ccc}
	\affaddr{{\uc3m}Universidad Carlos III de Madrid{\ }} & \affaddr{{\imdea}IMDEA Networks Institute{\ }} & \affaddr{{\nec}NEC Labs Europe{\ }} \\
	\affaddr{Avda. Universidad, 30} & \affaddr{Avda. Mar Mediterr\'{a}neo, 22} & \affaddr{Kurfürsten-Anlage, 36} \\
	\affaddr{28911 Legan\'{e}s, Spain} & \affaddr{28918 Legan\'{e}s, Spain} & \affaddr{69115 Heidelberg, Germany} \\
\end{tabular}
}

\maketitle

\begin{abstract}

Rate adaptation in 802.11 WLANs has received a lot of attention from the research community, with most of the proposals aiming at maximising throughput based on network conditions. Considering energy consumption, an implicit assumption is that optimality in throughput implies optimality in energy efficiency, but this assumption has been recently put into question. In this paper, we address via analysis and experimentation the relation between throughput performance and energy efficiency in multi-rate 802.11 scenarios. We demonstrate the trade-off between these performance figures, confirming that they may not be simultaneously optimised, and analyse their sensitivity towards the energy consumption parameters of the device. Our results provide the means to design novel rate adaptation schemes that takes energy consumption into account. 



\end{abstract}


\keywords{WLAN; 802.11; Rate Adaptation; Energy Efficiency}

\section{Introduction}
\label{sec:introduction}

In recent years, along with the exponential growth in mobile data applications and the corresponding traffic volume demand (see e.g. \cite{CVNI}), we have witnessed an increased attention towards ``green operation'' of  networks, which is required to support a sustainable growth of the communication infrastructures. For the case of wireless communications, there is the added motivation of a limited energy supply (i.e., batteries), which has triggered a relatively large amount of work on energy efficiency \cite{survey}. It turns out, though, that energy efficiency and performance do not necessarily come hand in hand, as some recent research has pointed out \cite{tradeoff, balancing}, and that a criterion may be required to set a proper balance between them. 

This paper is devoted to the problem of rate adaptation (RA) in 802.11 WLANs from the energy consumption's perspective. RA algorithms are responsible for selecting the most appropriate modulation and coding scheme (MCS) and transmission power (TXP) to use, given an estimation of the link conditions, and have received a vast amount of attention from the research community (see e.g. \cite{h-rca} and references therein). In general, the challenge lies in distinguishing between those loses due to collisions and those due to poor radio conditions, because they should trigger different reactions. In addition, the performance figure to optimise is commonly the throughput or a related one such as, e.g., the time required to deliver a frame.

It is generally assumed that optimality in terms of throughput also implies optimality in terms of energy efficiency. However, some recent work \cite{Li2012, khan2013} has shown that throughput maximisation does not result in energy efficiency maximisation, at least for 802.11n. However, we still lack a proper understanding of the causes behind this ``non-duality'', as it may be caused by the specific design of the algorithms studied, the extra consumption caused by the complexity of MIMO techniques, or any other reason. In fact, it could be an inherent trade-off given by the power consumption characteristics of 802.11 interfaces, and, if so, RA techniques should not be agnostic to this case.

This work tackles the latter question from a formal standpoint. A question which, to the best of the authors' knowledge, has never been addressed in the literature. For this purpose, and with the aim of isolating the variables of interest, we present a joint goodput and energy consumption model for single 802.11 spatial streams in the absence of interfering traffic. Packet losses occur due to poor channel conditions and RA can tune only two variables: MCS and TXP.

Building on this model, we provide the following contributions: ($i$)~we demonstrate through an extensive numerical evaluation that energy consumption and throughput performance are different optimisation objectives in 802.11, and not only an effect of MIMO or certain algorithms' suboptimalities; ($ii$)~we analyse the relative impact of each energy consumption component on the resulting performance of RA, which serves to identify the critical factors to consider for the design of RA algorithms, and illustrate that different hardware should employ different configurations; and ($iii$)~we experimentally validate our numerical results.

The rest of this paper is organised as follows. In Section~\ref{sec:models}, we develop the theoretical framework: a joint goodput-energy model built around separate previous models. In Section~\ref{sec:results}, we provide a detailed analysis of the trade-off between energy efficiency and maximum goodput, including a discussion of the role of the different energy parameters involved. In Section~\ref{sec:experiments}, we support our numerical analysis with experimental results. Finally, Section~\ref{sec:summary} summarises the paper. 

\section{Joint Goodput-Energy Model}\label{sec:models}

In this section, we develop a joint goodput-energy model for a single 802.11 spatial stream and the absence of interfering traffic. It is based on previous studies about goodput and energy consumption of wireless devices. As stated in the introduction, the aim of this model is the isolation of the relevant variables (MCS and TXP) to let us delve in the relationship between goodput and energy consumption optimality in the absence of other effects such as collisions or MIMO.

Beyond this primary intent, it is worth noting that these assumptions conform with real-world scenarios in the scope of recent trends in the IEEE 802.11 standard development, namely, the amendments 11ac and 11ad, where device-to-device communications (mainly through beamforming and MU-MIMO) are of paramount importance.

\subsection{Goodput Model}

We base our study on the work by Qiao \textit{et al.} \cite{Qiao2002}, which develops a robust goodput model that meets the established requirements. This model analyses the IEEE 802.11a Distributed Coordination Function (DCF) over the assumption of an AWGN (Additive White Gaussian Noise) channel without interfering traffic.

Let us briefly introduce the reader to the main concepts, essential to our analysis, of the goodput model by Qiao \textit{et al.}. Given a packet of length $l$ ready to be sent, a frame retry limit $n_\mathrm{max}$ and a set of channel conditions $\hat{s}=\{s_1, \ldots, s_{n_\mathrm{max}}\}$ and modulations $\hat{m}=\{m_1, \ldots, m_{n_\mathrm{max}}\}$ used during the potential transmission attempts, the \emph{expected effective goodput} $\mathcal{G}$ is modelled as the ratio between the expected delivered data payload and the expected transmission time as follows:
\begin{align}\label{goodput}
 \mathcal{G}(l, \hat{s}, \hat{m}) = \frac{\e{\mathrm{data}}}{\e{\mathcal{D}_\mathrm{data}}} = \frac{\Pr[\mathrm{succ} \mid l, \hat{s}, \hat{m}]\cdot l}{\e{\mathcal{D}_\mathrm{data}}}
\end{align}

\noindent where $\Pr[\mathrm{succ} \mid l, \hat{s}, \hat{m}]$ is the probability of successful transmission conditioned to $l, \hat{s}, \hat{m}$, given by Equation~(5) in \cite{Qiao2002}. The expected transmission time is defined as follows:
\begin{align}\label{Ddata}
 \e{\mathcal{D}_\mathrm{data}} = \left(1 - \Pr[\mathrm{succ} \mid l, \hat{s}, \hat{m}]\right) \cdot \mathcal{D}_{\mathrm{fail} \mid l, \hat{s}, \hat{m}} \\
 + \Pr[\mathrm{succ} \mid l, \hat{s}, \hat{m}] \cdot \mathcal{D}_{\mathrm{succ} \mid l, \hat{s}, \hat{m}} \nonumber
\end{align}

\noindent where 
\begin{align}\label{Dsucc}
 \mathcal{D}_{\mathrm{succ} \mid l, \hat{s}, \hat{m}} = &\sum_{n=1}^{n_\mathrm{max}} \Pr[n \mathrm{~succ} \mid l, \hat{s}, \hat{m}] \cdot \biggl\lbrace \sum_{i=2}^{n_\mathrm{max}} \left[\overline{T}_\mathrm{bkoff}(i)\right.\biggr. \nonumber\\
 &+ \left.T_\mathrm{data}(l, m_i) + \overline{\mathcal{D}}_\mathrm{wait}(i)\right] \nonumber\\
 &+ \overline{T}_\mathrm{bkoff}(1) + T_\mathrm{data}(l, m_1) + T_\mathrm{SIFS} \nonumber\\
 &+ \biggl.T_\mathrm{ACK}(m'_n) + T_\mathrm{DIFS} \biggr\rbrace
\end{align}

\noindent is the average duration of a successful transmission and
\begin{align}\label{Dfail}
 \mathcal{D}_{\mathrm{fail} \mid l, \hat{s}, \hat{m}} = &\sum_{i=1}^{n_\mathrm{max}} \left[\overline{T}_\mathrm{bkoff}(i)\right. \\
 &+ \left.T_\mathrm{data}(l, m_i) + \overline{\mathcal{D}}_\mathrm{wait}(i+1)\right] \nonumber
\end{align}

\noindent is the average time wasted during the $n_\mathrm{max}$ attempts when the transmission fails. 

$\Pr[n \mathrm{~succ} \mid l, \hat{s}, \hat{m}]$ is the probability of successful transmission at the $n$-th attempt conditioned to $l, \hat{s}, \hat{m}$, and $\overline{\mathcal{D}}_\mathrm{wait}(i)$ is the average waiting time before the $i$-th attempt. Their expressions are given by Equations~(7) and (8) in \cite{Qiao2002}. The transmission time ($T_\mathrm{data}$), ACK time ($T_\mathrm{ACK}$) and average backoff time ($\overline{T}_\mathrm{bkoff}$) are given by Equations~(1), (2) and (3) in \cite{Qiao2002}. Finally, $T_\mathrm{SIFS}$ and $T_\mathrm{DIFS}$ are 802.11a parameters, and they can be found also in Table~2 in \cite{Qiao2002}.

\subsection{Energy Consumption Model}

The selected energy model is our previous work of \cite{Serrano2014}, which has been further validated via ad-hoc circuitry and specialised hardware \cite{deseeding} and, to the best of our knowledge, stands as the most accurate energy model for 802.11 devices published so far, because it accounts not only the energy consumed by the wireless card, but the consumption of the whole device. While classical models focused on the wireless interface solely, this one demonstrates empirically that the energy consumed by the device itself cannot be neglected as a device-dependent constant contribution. Conversely, devices incur an energy cost derived from the frame processing, which may impact the relationship that we want to evaluate in this paper.

This model can be summarised as follows:
\begin{align}
 \overline{P} = \rho_\mathrm{id} + \sum_{i\in\mathrm{\{tx,rx\}}} \rho_i \tau_i + \sum_{i\in\mathrm{\{g,r\}}} \gamma_{\mathrm{x}i} \lambda_i
\end{align}

\noindent where $\rho_\mathrm{id}, \rho_\mathrm{tx}, \rho_\mathrm{rx}$ are the power consumed by the device in idle, transmission and reception states respectively; $\tau_\mathrm{tx}, \tau_\mathrm{rx}$ are the airtime percentages in transmission and reception; $\gamma_\mathrm{xg}, \gamma_\mathrm{xr}$ are the so called \emph{cross-factors}, a per-frame energy toll for generation and reception respectively; and $\lambda_\mathrm{g}, \lambda_\mathrm{r}$ are the frame generation and reception rates.

Therefore, the average power consumed $\overline{P}$ is a function of five device-dependent parameters ($\rho_i, \gamma_{\mathrm{x}i}$) and four traffic-dependent ones ($\tau_i, \lambda_i$).

\subsection{Energy Efficiency Analysis}

Putting together both models, we are now in a position to build a joint goodput-energy model for 802.11a DCF. Let us consider the average durations \eqref{Dsucc} and \eqref{Dfail}. Based on their expressions, we multiply the idle time ($\overline{\mathcal{D}}_\mathrm{wait}$, $\overline{T}_\mathrm{bkoff}$, $T_\mathrm{SIFS}$, $T_\mathrm{DIFS}$) by $\rho_\mathrm{id}$, the transmission time ($T_\mathrm{data}$) by $\rho_\mathrm{tx}$, and the reception time ($T_\mathrm{ACK}$) by $\rho_\mathrm{rx}$. The resulting expressions are the average energy consumed in a successful transmission $\mathcal{E}_{\mathrm{succ} \mid l, \hat{s}, \hat{m}}$ and the average energy wasted when a transmission fails $\mathcal{E}_{\mathrm{fail} \mid l, \hat{s}, \hat{m}}$:
\begin{align}
 \mathcal{E}_{\mathrm{succ} \mid l, \hat{s}, \hat{m}} = &\sum_{n=1}^{n_\mathrm{max}} \Pr[n \mathrm{~succ} \mid l, \hat{s}, \hat{m}] \cdot \biggl\lbrace \sum_{i=2}^{n_\mathrm{max}} \left[\rho_\mathrm{id}\overline{T}_\mathrm{bkoff}(i)\right.\biggr. \nonumber\\
 &+ \left.\rho_\mathrm{tx}T_\mathrm{data}(l, m_i) + \rho_\mathrm{id}\overline{\mathcal{D}}_\mathrm{wait}(i)\right] \nonumber\\
 &+ \rho_\mathrm{id}\overline{T}_\mathrm{bkoff}(1) + \rho_\mathrm{tx}T_\mathrm{data}(l, m_1) + \rho_\mathrm{id}T_\mathrm{SIFS} \nonumber\\
 &+ \biggl.\rho_\mathrm{rx}T_\mathrm{ACK}(m'_n) + \rho_\mathrm{id}T_\mathrm{DIFS} \biggr\rbrace
\end{align}
\begin{align}
 \mathcal{E}_{\mathrm{fail} \mid l, \hat{s}, \hat{m}} = &\sum_{i=1}^{n_\mathrm{max}} \left[\rho_\mathrm{id}\overline{T}_\mathrm{bkoff}(i)\right. \\
 &+ \left.\rho_\mathrm{tx}T_\mathrm{data}(l, m_i) + \rho_\mathrm{id}\overline{\mathcal{D}}_\mathrm{wait}(i+1)\right] \nonumber
\end{align}


Then, by analogy with \eqref{Ddata}, the \emph{expected energy consumed per frame transmitted}, $\e{\mathcal{E}_\mathrm{data}}$, can be written as follows:
\begin{align}\label{energyperframe}
 \e{\mathcal{E}_\mathrm{data}} = \gamma_\mathrm{xg} + \left(1 - \Pr[\mathrm{succ} \mid l, \hat{s}, \hat{m}]\right) \cdot \mathcal{E}_{\mathrm{fail} \mid l, \hat{s}, \hat{m}} \\
 + \Pr[\mathrm{succ} \mid l, \hat{s}, \hat{m}] \cdot \mathcal{E}_{\mathrm{succ} \mid l, \hat{s}, \hat{m}} \nonumber
\end{align}

It is noteworthy that the receiving cross-factor does not appear in this expression because ACKs (acknowledgements) are processed in the network card exclusively, and thus its processing toll is negligible.

Finally, we define the \emph{expected effective energy efficiency} $\mu$ as the ratio between the expected delivered data payload and the expected energy consumed per frame, which can be expressed in \emph{bits per Joule} (bpJ):
\begin{align}\label{efficiency}
 \mu(l, \hat{s}, \hat{m}) = \frac{\e{\mathrm{data}}}{\e{\mathcal{E}_\mathrm{data}}}
\end{align}

\section{Numerical Results}\label{sec:results}


Building on the joint model presented in the previous section, here we explore the relationship between optimal goodput and energy efficiency in 802.11a. More specifically, our objective is to understand the behaviour of the energy efficiency of a single spatial stream as the MCS and TXP change following our model to meet the optimal goodput.

\subsection{Optimal Goodput}

We note that the main goal of RA, generally, is to maximise the effective goodput that a station can achieve by varying the parameters of the interface. In terms of the model discussed in the previous section, a rate adaptation algorithm would aspire to fit the following curve:
\begin{align}\label{maxgoodput}
 \max{\mathcal{G}(l, \hat{s}, \hat{m})}
\end{align}

\begin{figure}[t]
	\centering
	\includegraphics[width=\linewidth]{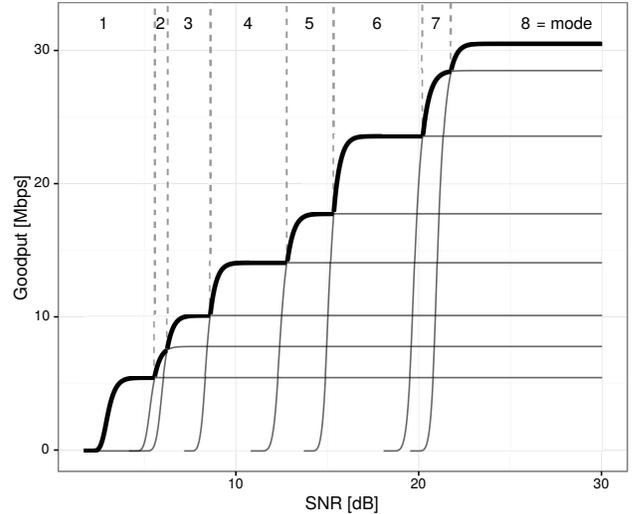}
	\caption{Optimal goodput (bold envelope) as a function of SNR.}
	\label{fig:maxgoodput}
\end{figure}

We provide the numerical results for this goodput maximisation problem in Fig.~\ref{fig:maxgoodput}, which are in good agreement with those obtained in \cite{Qiao2002}. For the sake of simplicity but without loss of generality we fix $l=1500$ octets and $n_\mathrm{max}=7$ retries, and assume that the channel conditions and the transmission strategy are constant across retries ($\hat{s}=\{s_1, \ldots, s_1\}$ and $\hat{m}=\{m_1, \ldots, m_1\}$).

Fig.~\ref{fig:maxgoodput} illustrates which mode (see Table~\ref{tab:modes}) is optimal in terms of goodput, given an SNR level. We next address the question of whether this optimisation is aligned with energy efficiency maximisation.

\begin{table}[t]
	\renewcommand{\arraystretch}{1.3}
	\footnotesize
	\caption{Modes of the IEEE 802.11a PHY}
	\label{tab:modes}
	\centering
	
	\begin{tabular}{r|rrrrrrrr}
	 \hline 
	 Mode Index & 1 & 2 & 3 & 4 & 5 & 6 & 7 & 8 \\ 
	 \hline 
	 MCS (Mbps) & 6 & 9 & 12 & 18 & 24 & 36 & 48 & 54 \\ 
	 \hline 
	 \end{tabular}  
\end{table}

\subsection{Extension of the Energy Parametrisation}

\begin{table*}[t]
	\renewcommand{\arraystretch}{1.3}
	\footnotesize
	\caption{Linear Regressions}
	\label{tab:regressions_tx}
	\centering
	\begin{tabular}{r|lll|c|ll|c}
		\hline 
		\multirow{2}{*}{Device} & \multicolumn{4}{c|}{$\rho_\mathrm{tx}$ model estimates ($\alpha_i$)} & \multicolumn{3}{c}{$\rho_\mathrm{rx}$ model estimates ($\beta_i$)} \\
		\cline{2-8}
		& (Intercept) [W] & MCS [Mbps] & TXP [mW] & adj. $r^2$ & (Intercept) [W] & MCS [Mbps] & adj. $r^2$ \\
		\hline 
		HTC Legend & 0.354(14) & 0.0052(3) & 0.021(3) & 0.97 & 0.013(3) & 0.00643(11) & $>$0.99 \\
		Linksys WRT54G & 0.540(12) & 0.0028(2) & 0.075(3) & 0.98 & 0.14(2) & 0.0130(7) & 0.96 \\
		Raspberry Pi & 0.478(19) & 0.0008(4) & 0.044(5) & 0.88 & -0.0062(14) & 0.00146(5) & 0.98 \\
		Galaxy Note 10.1 & 0.572(4) & 0.0017(1) & 0.0105(9) & 0.98 & 0.0409(10) & 0.00173(4) & 0.99 \\
		Soekris net4826-48 & 0.17(3) & 0.0170(6) & 0.101(7) & 0.99 & 0.010(8) &0.0237(3) & $>$0.99 \\
		\hline 
	\end{tabular}
\end{table*}

\begin{figure*}[t]
	\centering
	\csubfloat[$\rho_\mathrm{tx}$ fit as a function of MCS and TXP.]{
		\includegraphics[width=\linewidth]{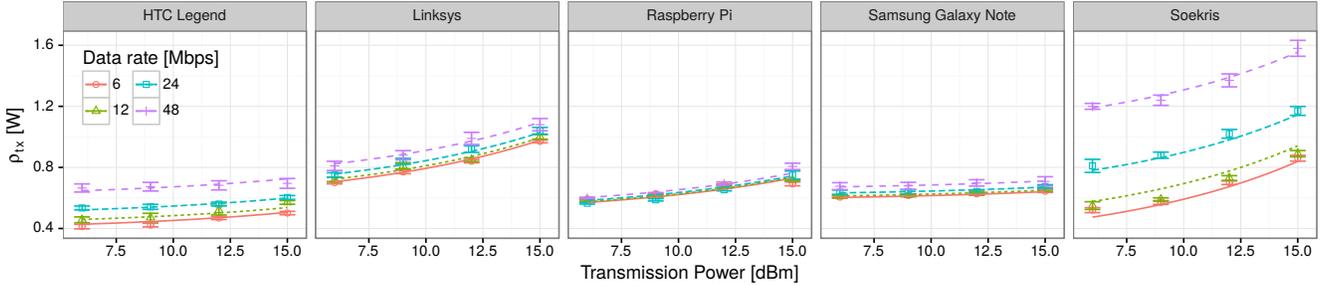}
		\label{fig:rho_tx}
	}\\
	\csubfloat[$\rho_\mathrm{rx}$ fit as a function of MCS.]{
		\includegraphics[width=\linewidth]{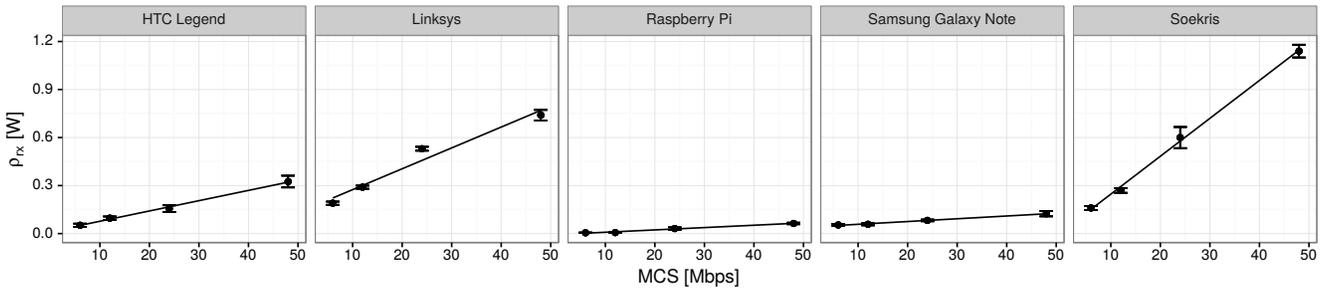}
		\label{fig:rho_rx}
	}
	\caption{Linear regressions.}
\end{figure*}

The next step is to delve into the energy consumption of wireless devices. \cite{Serrano2014} provides real measurements for five devices: three AP-like platforms (Linksys WRT54G, Raspberry Pi and Soekris net4826-48) and two hand-held devices (HTC Legend and Samsung Galaxy Note 10.1). Two of the four parameters needed are constant ($\rho_\mathrm{id}, \gamma_\mathrm{xg}$), and the other two ($\rho_\mathrm{tx}, \rho_\mathrm{rx}$) depend on the MCS and the TXP used. However, the characterisation done in \cite{Serrano2014} is performed for a subset of the MCS and TXP available, so we next detail how we extend the model to account for a larger set of operation parameters.

A detailed analysis of the numerical figures presented in \cite{Serrano2014} suggests that $\rho_\mathrm{rx}$ depends linearly on the MCS, and that $\rho_\mathrm{tx}$ depends linearly on the MCS and the TXP (in mW). Based on these observations, we define the following linear models:
\begin{align}
 \rho_\mathrm{tx} &= \alpha_0 + \alpha_1\cdot\mathrm{MCS} + \alpha_2\cdot\mathrm{TXP} \\
 \rho_\mathrm{rx} &= \beta_0 + \beta_1\cdot\mathrm{MCS}
\end{align}

The models are fed with the data reported in \cite{Serrano2014}, and the resulting fitting is illustrated in Figs.~\ref{fig:rho_tx} and \ref{fig:rho_rx}, while Table~\ref{tab:regressions_tx} collects the model estimates for each device (with errors between parentheses), as well as the adjusted r-squared. Since these linear models show a very good fit, they support the generation of synthetic data for the different MCS and TXP required.

\subsection{Energy Consumption}

\begin{figure*}[t]
	\centering
	\includegraphics[width=\linewidth]{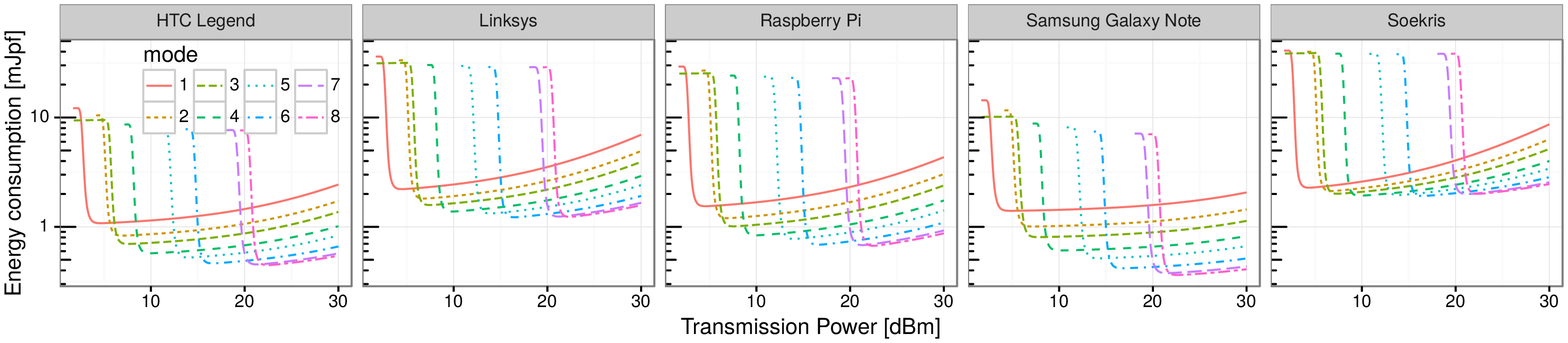}
	\caption{Expected energy consumption per frame in \emph{millijoules per frame} (mJpf) under fixed channel conditions.}
	\label{fig:consumption}
\end{figure*}

To compute the energy consumption using the above parametrisation, first we have to define the assumptions for the considered scenario. We assume for simplicity a device-to-device communication, with fixed and reciprocal channel conditions during a sufficient period of time (i.e., low or no mobility). As we have discussed before, our primary goal is to isolate MCS and TXP as variables of interest, but we must not forget that these are also reasonable assumptions in scenarios targeted by recent 802.11 standard developments (11ac, 11ad).

For instance, given channel state information from a receiver, the transmitter may decide to increase the TXP in order to increase the receiver's SNR (and thus the expected goodput), or to decrease it if the channel quality is high enough. Although the actual relationship between TXP and SNR depends on the specific channel model (e.g., distance, obstacles, noise), without loss of generality, we choose a noise floor of $N=-85$~dBm in an office scenario with a link distance of $d=18$~m in order to explore numerically the whole range of SNR while using reasonable values of TXP. The ITU model for indoor attenuation \cite{iturp1238-2015} gives a path loss of $L\approx 85$~dBm. Then, we can use \eqref{energyperframe} to obtain the expected energy consumed per frame and MCS mode, with TXP being directly related to the SNR level.

The results are reported in Fig.~\ref{fig:consumption}. As the figure illustrates, consumption first falls abruptly as the TXP increases for all modes, which is caused when the SNR reaches a sharp threshold level such that the number of retransmissions changes from 6 to 0 (i.e., no frame is discarded). From this threshold on, the consumption increases with TXP because, although the number of retransmissions is 0, the wireless interface consumes more power. We note that the actual value of the TXP when the consumption drops depends on the specifics of the scenario considered, but the qualitative conclusions hold for a variety of scenarios.

\subsection{Energy Efficiency vs. Optimal Goodput}

\begin{figure}[t]
	\centering
	\includegraphics[width=\linewidth]{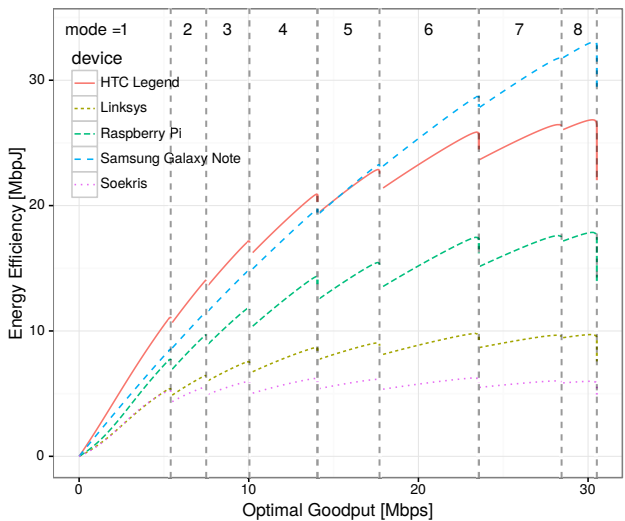}
	\caption{Energy Efficiency vs. Optimal Goodput under fixed channel conditions.}
	\label{fig:efficiency-goodput}
\end{figure}

We can finally merge previous numerical analyses and confront energy efficiency, given by \eqref{efficiency}, and optimal goodput, given by \eqref{maxgoodput}, for all devices and under the aforementioned assumptions. To this aim, we plot in the same figure the energy efficiency for the configuration that maximises goodput given an SNR value vs. the obtained goodput, with the results being depicted in Fig.~\ref{fig:efficiency-goodput}. We next discuss the main findings from the figure. 

First of all, we can see that the energy efficiency grows sub-linearly with the optimal goodput (the optimal goodput for each SNR value) in all cases. We may distinguish three different cases in terms of energy efficiency: high (Samsung Galaxy Note and HTC Legend), medium (Raspberry Pi) and low energy efficiency (Linksys and Soekris). Furthermore, for the case of the Soekris, we note that the ``central modes'' (namely, 4 and 5) are more efficient in their optimal region than the subsequent ones.

Another finding (more relevant perhaps) is that it becomes evident that increasing the goodput does not always improve the energy efficiency: there are more or less drastic leaps, depending on the device, between mode transitions. From the transmitter point of view, in the described scenario, this can be read as follows: we may increase the TXP to increase the SNR, but if the optimal goodput entails a mode transition, the energy efficiency may be affected.

As a conclusion, we have demonstrated that optimal goodput and energy efficiency do not go hand in hand, even in a single spatial stream, in 802.11. There is a trade-off in some circumstances that current rate adaptation algorithms cannot take into account, as they are oblivious to the energy consumption characteristic of the device. 

\subsection{Sensitivity to Energy Parameter Scaling}\label{sec:param-scaling}

We next explore how the different energy parameters affect the energy efficiency vs. optimal goodput relationship. For this purpose, we selected the Raspberry Pi curve from Fig.~\ref{fig:efficiency-goodput} (results are analogous with the other devices) and we scale up and down, and one at a time, the four energy parameters $\rho_\mathrm{id}$, $\rho_\mathrm{tx}$, $\rho_\mathrm{rx}$, and $\gamma_\mathrm{xg}$. The scaling up and down is done by multiplying and dividing by 3, respectively, the numerical value of the considered parameter. One of the first results is that the impact of $\rho_\mathrm{rx}$ is negligible, a result somehow expected as the cost of receiving the ACK is practically zero. From this point on, we do not consider further this parameter.

\begin{figure}[t]
	\centering
	\csubfloat[Overall effect.]{
		\includegraphics[width=\linewidth]{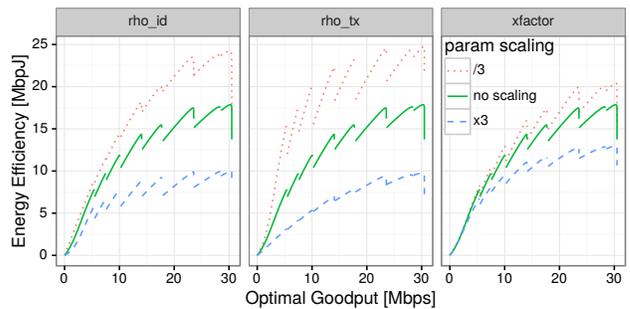}
		\label{fig:param-scaling1}
	}\\
	\csubfloat[Impact on mode transitions.]{
		\includegraphics[width=\linewidth]{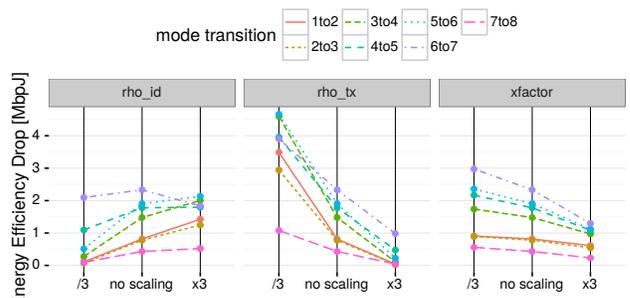}
		\label{fig:param-scaling2}
	}
	\caption{Impact of energy parameter scaling on the energy efficiency.}
\end{figure}

We show in Fig.~\ref{fig:param-scaling1} the overall effect of this parameter scaling. The solid line represents the base case with no scaling (same curve as in Fig.~\ref{fig:efficiency-goodput}), and in dashed and dotted lines the corresponding parameter was multiplied or divided by a factor of 3, respectively. As expected, larger parameters contribute to lower the overall energy efficiency. However, the impact on the energy efficiency drops between mode transitions is far from being obvious, as in some cases transitions are more subtle while in others they become more abrupt.

To delve into these transitions, we illustrate in Fig.~\ref{fig:param-scaling2} the ``drop'' in energy efficiency when changing between modes. As it can be seen, the cross-factor is the less sensitive parameter of the three, because its overall effect is limited and, more importantly, it scales all the leaps between mode transitions homogeneously. This means that a higher or lower cross-factor, which resides almost entirely in the device and not in the wireless card, does not alter the energy efficiency vs. optimal goodput relationship (note that this parameter results in a constant term in \eqref{energyperframe}). Thus, the cross-factor is not relevant from the RA point of view, and energy-aware RA algorithms can be implemented by leveraging energy parameters local to the wireless card.

On the other hand, $\rho_\mathrm{id}$ and $\rho_\mathrm{tx}$ have a larger overall effect, plus an inhomogeneous and, in general, opposite impact on mode transitions. While a larger $\rho_\mathrm{id}$ contributes to larger leaps, for the case of $\rho_\mathrm{tx}$, the larger energy efficiency drops occur with smaller values of that parameter. Still, the reason behind this behavior is the same for both cases: the wireless card spends more time in idle (and less time transmitting) when a transition to the next mode occurs, which has a higher data rate.

This effect is also evident if we compare the Samsung Galaxy Note and the HTC Legend curves in Fig.~\ref{fig:efficiency-goodput}. Both devices have $\rho_\mathrm{id}$ and $\rho_\mathrm{tx}$ in the same order of magnitude, but the HTC Legend has a larger $\rho_\mathrm{id}$ and a smaller $\rho_\mathrm{tx}$. The combined outcome is a more dramatic sub-linear behaviour and an increased energy efficiency drop between mode transitions.

\section{Experimental Validation}\label{sec:experiments}

This section is devoted to experimentally validate the results from the numerical analysis and, therefore, the resulting conclusions. To this aim, we describe our experimental setup and the validation procedure, first specifying the methodology and then the results achieved.

\subsection{Experimental Setup}

\begin{figure}[t]
	\centering
	\includegraphics[width=\linewidth]{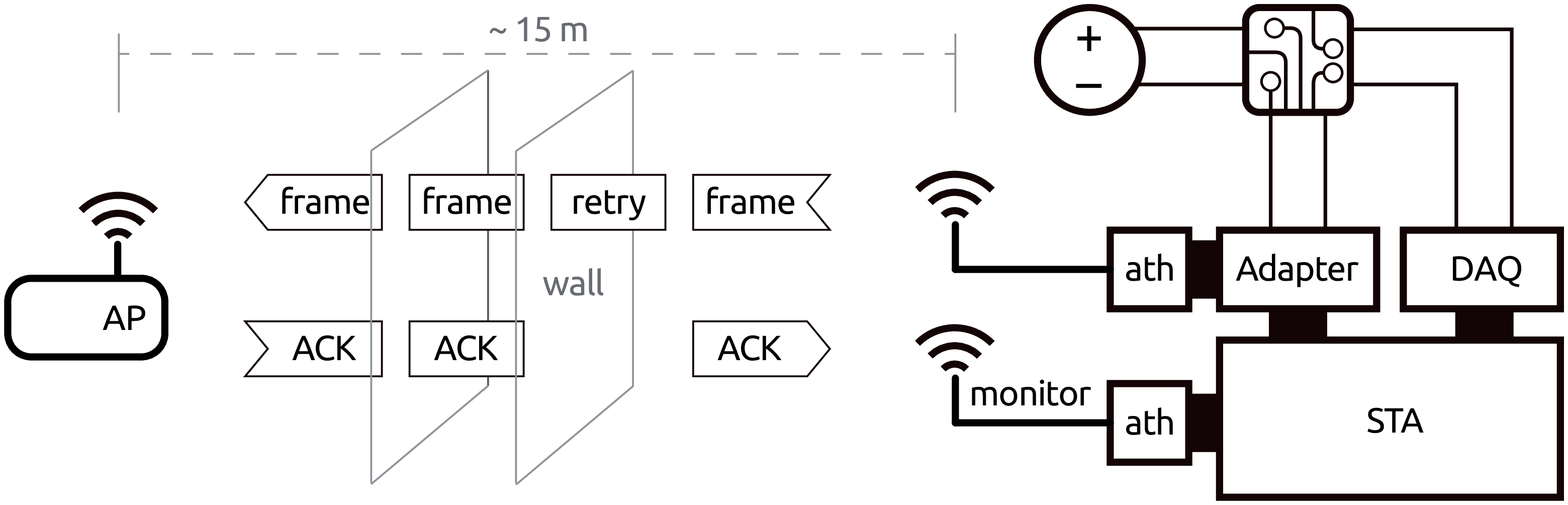}
	\caption{Experimental setup.}
	\label{fig:testbed}
\end{figure}

We deployed the testbed illustrated in Fig.~\ref{fig:testbed}, which consists of a station (STA) transmitting evenly-spaced maximum-sized UDP packets to an access point (AP). The AP is an x86-based Alix6f2 board with a Mini PCI Qualcomm Atheros AR9220 wireless network adapter, running Voyage Linux with kernel version 3.16.7 and the \texttt{ath9k} driver. The STA is a desktop PC with a Mini PCI Express Qualcomm Atheros QCA9880 wireless network adapter, running Fedora Linux 23 with kernel version 4.2.5 and the \texttt{ath10k} driver. We also installed at the STA a Mini PCI Qualcomm Atheros AR9220 wireless network adapter to monitor the wireless channel.

The QCA9880 card is connected to the PC through a \emph{x1 PCI Express to Mini PCI Express} adapter from Amfeltec. This adapter connects the PCI bus' data channels to the host and provides an ATX port so that the wireless card can be supplied by an external power source. The power supply is a Keithley 2304A DC Power Supply, and it powers the wireless card through an \emph{ad-hoc} measurement circuit that extracts the voltage and converts the current with a high-precision sensing resistor and amplifier. These signals are measured using a National Instruments PCI-6289 multifunction data acquisition (DAQ) device, which is also connected to the STA. Thanks to this configuration, the STA can simultaneously measure the instant power consumed by the QCA9880 card\footnote{Following the discussion on Section~\ref{sec:param-scaling}, the device's cross-factor is not involved in the trade-off, thus we will expect to reproduce it by measuring the wireless interface alone.} and the goodput achieved.

As the figure illustrates, the STA is located in an office space and the AP is placed in a laboratory 15~m away, and transmitted frames have to transverse two thin brick walls. The wireless card uses only one antenna and a practically-empty channel in the 5-GHz band. Throughout the experiments, the STA is constantly backlogged with data to send to the AP, and measures the throughput obtained by counting the number of received acknowledgements (ACKs).

\subsection{Methodology and Results}

\begin{figure}[t]
	\centering
	\includegraphics[width=\linewidth]{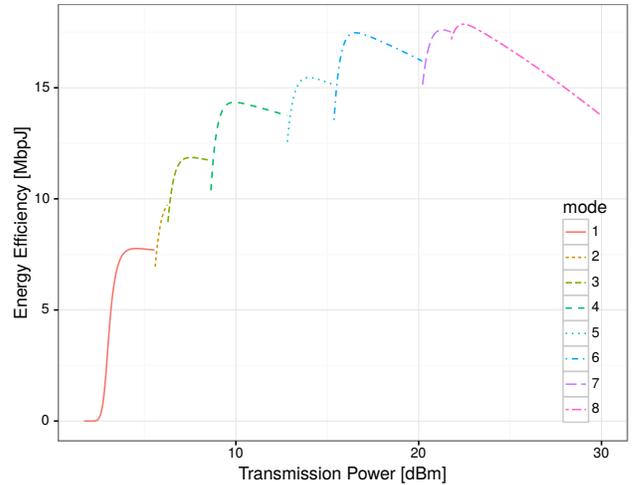}
	\caption{Energy Efficiency vs. Transmission Power under fixed channel conditions for the Raspberry Pi case.}
	\label{fig:efficiency-txp}
\end{figure}

In order to validate our results, our aim is to replicate the qualitative behaviour of Fig.~\ref{fig:efficiency-goodput}, in which there are energy efficiency ``drops'' as the optimal goodput increases. However, in our experimental setting, channel conditions are far from steady, which introduces a notable variability in the results as it affects both the $x$-axis (goodput) and the $y$-axis (energy efficiency). To reduce the impact of this variability, we decided to change the variable in the $x$-axis from the optimal goodput to the transmission power --a variable that is directly configured in the wireless card. In this way, the qualitative behaviour to replicate is the one illustrated in Fig.~\ref{fig:efficiency-txp}, where we can still identify the performance ``drops'' causing the loss in energy efficiency.



\begin{figure}[t]
	\centering
	\includegraphics[width=\linewidth]{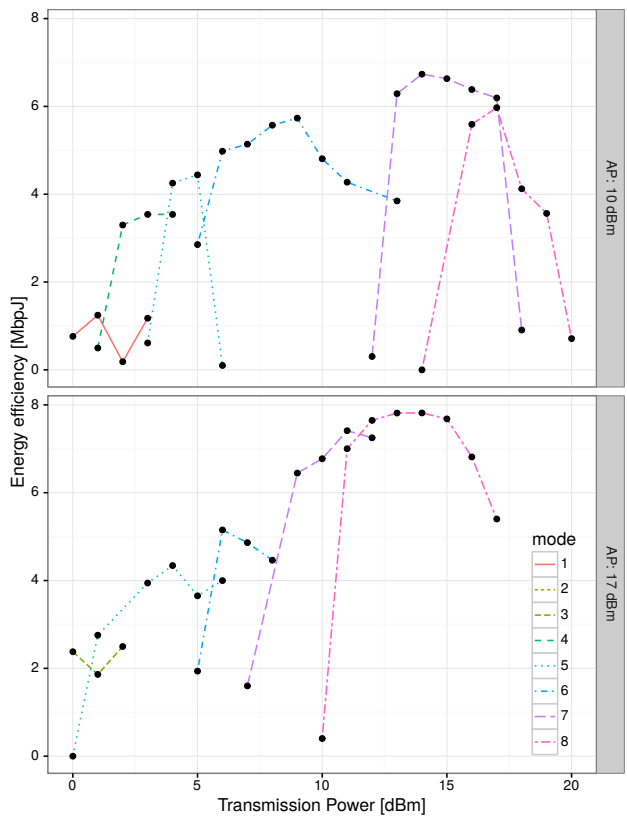}
	\caption{Experimental study of Fig.~\ref{fig:efficiency-txp} for two AP configurations.}
	\label{fig:efficiency-txp-exp}
\end{figure}

Building on this figure, we perform a sweep through all available combinations of MCS (see Table~\ref{tab:modes}) and TXP\footnote{The model explores a range between 0 and 30 dBm to get the big picture, but this particular wireless card only allows us to sweep from 0 to 20 dBm.}. Furthermore, we also tested two different configurations of the AP's TXP at different times of the day, to confirm that this qualitative behaviour is still present under different channel conditions. For each configuration, we performed 2-second experiments in which we measure the total bytes successfully sent and the energy consumed by the QCA9880 card with sub-microsecond precision. From this data, the energy efficiency is computed, with the results depicted in Fig.~\ref{fig:efficiency-txp-exp} (each figure corresponds to a different TXP value configured at the AP). 



In the figure, each line type represents the STA's mode that achieved the highest goodput for each TXP interval, therefore in some cases low modes do not appear because a higher mode achieved a higher goodput. Despite the inherent experimental difficulties, namely, the low granularity of 1-dBm steps and the random variability of the channel, the experimental results validate the analytical ones, as the qualitative behaviour of both figures follows the one illustrated in Fig.~\ref{fig:efficiency-txp}. In particular, the performance ``drops'' of each dominant mode can be clearly observed (especially for the 36, 48 and 54 Mbps MCSs) despite the variability in the results.

\section{Summary and Future Work}\label{sec:summary}

In this paper, we have revisited 802.11 rate adaptation by taking energy consumption into account. While some previous studies pointed out that MIMO rate adaptation is not energy efficient, we have demonstrated through numerical analysis that, even for single spatial streams without interfering traffic, energy consumption and throughput performance are different optimisation objectives. Furthermore, we have validated our results via experimentation.

Our findings show that this trade-off emerges at certain ``mode transitions'' when maximising the goodput, suggesting that small goodput degradations may lead to energy efficiency gains. 
Moreover, our analyses have showed that these trade-offs arise as a consequence of the power consumption behaviour of wireless cards and does not depend on the energy consumed in the rest of the device. In this way, energy-aware rate adaptation may be achieved building on information local to the wireless interface. Still, to develop energy-aware rate adaptation algorithms, further research is needed to understand how the findings of this work can be leveraged in suboptimal conditions, and how other effects, such as collisions and MIMO, affect the established trade-off.

\section{Acknowledgements}

This work has been performed in the framework of the H2020-ICT-2014-2 projects 5GNORMA (grant agreement no. 671584) and Flex5Gware (grant agreement no. 671563). The authors would like to acknowledge the contributions of their colleagues. This information reflects the consortium's view, but the consortium is not liable for any use that may be made of any of the information contained therein.

\bibliographystyle{abbrv}
\bibliography{mswim053-ucar}
\balance
\end{document}